\documentclass[aps,pre,floats,twocolumn]{revtex4}
\usepackage{graphicx}
\begin{document}
\title{Dynamics of conduction blocks in a model of paced cardiac tissue}
\author{Herv\'e Henry}
\altaffiliation[Currently at ]{Laboratoire de Physique de la Mati\`ere Condens\'ee, Ecole Polytechnique, 91128 Palaiseau, France}
\author{Wouter-Jan Rappel}              
\affiliation{Center for Theoretical Biological Physics\\ University
of California San Diego\\ 9500 Gilman Dr \\La Jolla CA, USA}
\date{\today}

\begin{abstract}
We study numerically the dynamics of conduction blocks
using a detailed electrophysiological model. 
We find that this dynamics depends critically on the size of the paced
region.
Small pacing regions lead to stationary conduction blocks while
larger pacing regions can lead to conduction blocks that 
travel periodically towards the pacing region. 
We show that this size-dependence dynamics can lead to a novel 
arrhythmogenic mechanism. 
Furthermore, we show that the essential phenomena can be 
captured in a much simpler coupled-map model.
\end{abstract}
\maketitle

\section{Introduction}
The coupling of the electrical excitation to the contractile
forces in the heart is essential to the 
blood supply to the body. 
Abnormal electrical activity, in particular arrhythmias, can have
dire consequences. The most serious of all, 
ventricular fibrillation (VF), is characterized by 
disordered electrical activity within the ventricles
and leads to death within minutes.

Unfortunately, the precise cause for the onset and maintenance
of VF remains elusive. 
It is believed, however, that electrical wave reentry 
plays an important role. 
Reentry is initiated when an electrical wave 
is locally blocked, leading to a broken wave front that 
can re-excite the tissue
behind the conduction block.
There are several ways to create a conduction block. 
Perhaps the most intuitive one is an
anatomical tissue heterogeneity through which the 
electrical wave fails to propagate\cite{Akaetal02, HenRap04}.
Although these heterogeneities exist and play a role in the 
initiation of VF, recently a new mechanism for conduction block
in perfectly {\it homogeneous} tissue has been described in 
models \cite{Quetal00a, Watetal01, Foxetald02} and experiments
\cite{Pasetal99, Foxetala02}.
Key contributor to this {\it dynamical} heterogeneity is 
electrical alternans which is characterized by 
a beat to beat oscillation in the action potential duration (APD,
defined below) 
under rapid pacing conditions.
Clinically, the occurrence of T wave alternans in ECGs has been 
associated with sudden cardiac arrest \cite{Rosetal94, Estetal97}. 
In isolated cells, the onset of alternans can be 
determined from the restitution curve which relates
the APD to the diastolic interval (DI),
the time interval between two successive action potential.
If this curve has a slope larger than one, it is 
easy to see that a period doubling bifurcation develops, 
resulting in an APD that is alternatingly 
short and long \cite{NolDah68,Gueetal89}.  
This simple vision, however, is most likely not complete as
memory effects \cite{Foxetalc02} and calcium cycling effects
\cite{Shietal03,Pruetal04} have been shown to play a role in the 
mechanism leading to alternans.

In the case of spatially extended systems, the 
possible dynamical behavior becomes more complicated. 
The conduction velocity also depends on the 
DI, which can 
lead to a spatial modulation  of the alternans, called
discordant alternans.
\cite{Quetal00a, Watetal01}. During 
discordant alternans, 
the APD is following a long-short-long pattern in one region
of the cable and  a short-long-short pattern in another. 
Separating these regions are nodes where  the APD is constant. They 
can be either stationary or traveling. 
In addition, during discordant alternans,  
the amplitude of the alternans (the difference between 
APDs in subsequent beats) can grow when moving away from the pacing site.
Hence, at a critical distance, 
provided the amplitude of alternans is large enough, 
the tissue can no longer support 
a traveling wave for each excitation and a diffusive conduction block\cite{Bik02} 
develops: 
a wave will propagate over a given distance along the cable and 
will then come to a stop. 

Here, we study the dynamics of the conduction blocks 
using a detailed electrophysiological model and
a pacing domain of variable size.
We present a quantitative and qualitative
description of the observed patterns and present results from
numerical experiments in cables and sheets of cardiac tissue.
Our main results are: (i) in addition to a stationary conduction block, 
rapid pacing can lead to 
periodic patterns of moving conduction blocks, (ii) the 
type of conduction block is critically dependent on the 
size of the pacing region, and (iii) these effects combined
can lead to a novel mechanism for reentry. 
Furthermore, we show that a simple coupled-map model 
gives qualitatively similar results.

\section{Modeling of a cable of cardiac cells and terminology}
To describe the electrophysiological properties of cardiac
tissue we used the reaction-diffusion equation: 
\begin{equation}
{\partial_t V}= D\nabla^2 V -(\Sigma_{ion} I_{ion}+I_{stim})/{C}
\label{oned}
\end{equation}
where $V$ is the transmembrane potential, 
$C=1\ \mu \rm{F}\, cm^{-2}$ is the membrane capacitance, 
$D \nabla^2, \mbox{ with } D=0.001 cm^2 s^{-1}$, expresses the inter-cellular
coupling, $I_{ion}$ represents
the different trans-membrane currents and $I_{stim}$ is the 
applied pacing current.  
The  ionic currents in $I_{ion}$
are governed by nonlinear evolution equations coupled
to $V$ and here we have chosen to use a
modified version of the
Luo-Rudy model \cite{LuoRud91} to describe these currents. 
A detailed description of this model and 
our modifications are given in appendix \ref{LR1Dyn}. 
The integration scheme we used  was forward Euler with space and time 
discretizations of $\delta x=0.025$ cm and $\delta t=0.02$ ms, respectively.
Finally, the space constant was determined to be 0.06 cm.

The cable was paced 
at one end with a period $T$ through  a 1 ms long 
constant current stimulus of $I_{stim}=-80 \mu A/cm^2$ \footnote{the threshold stimulus is approximately $-30 \mu A/cm^2$ when pacing  10 cells of the cable or when pacing a single cell and goes up to approximately  $-45 \mu A/cm^2$ when pacing a single cell of the cable.}
The pacing domain consisted of  $n$ gridpoints,
where $n$  was varied between 1 and 20, thus corresponding to a 
pacing domain ranging in size from 0.025 cm to 0.5 cm
\footnote{In order to check that our results are not due to a 
change in the total stimulation current  
we performed additional simulations. In these simulations, 
we kept the total stimulation current ($n I_{stim}$) constant by choosing 
$I_{stim}=- 800/n \ \mu A/cm^2$.
This change in stimulation protocol did not significantly alter
our results; the dynamics of conduction blocks away from the pacing
site was unchanged while conduction blocks at the pacing site occurred 
for slightly smaller  
values of $n$ or slightly smaller pacing period $T$. 
Hence, the results we report here can not be attributed to a 
change in total stimulation amplitude as in \cite{Yehia99}.}
As we will see below, changing the number of stimulated grid points can 
have a dramatic effect.
On the other hand, changing 
the size of the simulation cable from 10 cm, the length used here,
to 5 cm  did not reveal any qualitative differences. 
This, along with the fact that the propagation speed from endocardium
to epicardium can be much smaller (about 17 cm/s)\cite{Preetal95} than the one used here (about 51 cm/s), implies that
the observations reported here might be relevant to human
ventricles.
Finally, 
additional simulations performed using either a stronger stimulus or a 
longer stimulus showed little effect. 

\begin{figure}
\includegraphics[width=0.35\textwidth]{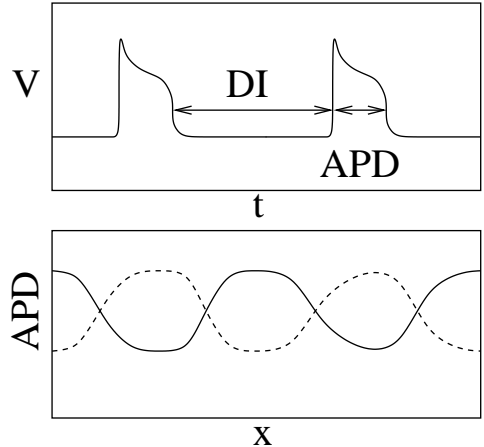}
\caption{top: Schematic time course of the transmembrane potential 
at a point during normal cardiac rhythm (i.e. without alternans). 
The diastolic interval (DI) is the period of time during which 
the transmembrane potential is below a given value and the 
action potential duration  (APD) is the period of time during which the action
potential is over this threshold value.  
Bottom: Schematic drawing of the APD along the cable during an even beat  
(solid line) and an odd beat (dashed line) 
in the discordant alternans regime. 
In the case of normal cardiac rhythm (i.e. without alternans), 
both lines would be the same and the APD would be constant in space 
(except at the pacing site (resp. the end of the cable) 
where boundary effects tend to lengthen (resp. shorten) the APD).
In the case of concordant alternans, the APD during successive
beats would be different but constant along the cable (apart from
boundary effects).}
\end{figure}
 
We define the beginning of an action potential at position $x$ as the moment 
when $V(x,t)$ crosses a threshold value $V_{th}$ from below.
Similarly, the end of the action potential is defined as the moment when 
$V(x,t)$ decreases below $V_{th}$. The APD is then defined as the time difference between these two 
events (see Fig. 1).
Then, the DI is defined as the 
time between the end of one action potential and the beginning of 
the next (see Fig. 1).
A conduction block for the $m^{th}$ stimulus occurs at the $i^{th}$ 
grid point when an action potential was elicited at the $i^{th}-1$ 
grid point by the $m^{th}$ stimulus but not at the $i^{th}$ one. 
$V_{th}$ was chosen to be equal to -60 mV and changes in the value of 
$V_{th}$ (between -80 mV and -50 mV) did not affect results significantly. 
In addition, it is useful to define a waveback and a wavefront. 
A wavefront is the boundary between a region at rest and an excited region 
when the latter region is invading the former 
while the waveback is the boundary when the former is invading the latter.  

The pacing periods used here ($\approx 160$ ms) are short
compared to the normal pacing period in humans. 
Of course, this pacing period is dictated by the
the  choice of the electrophysiological model used here. 

\section{Results in a cable \label{results}}
\begin{figure}
\includegraphics[width=0.5\textwidth]{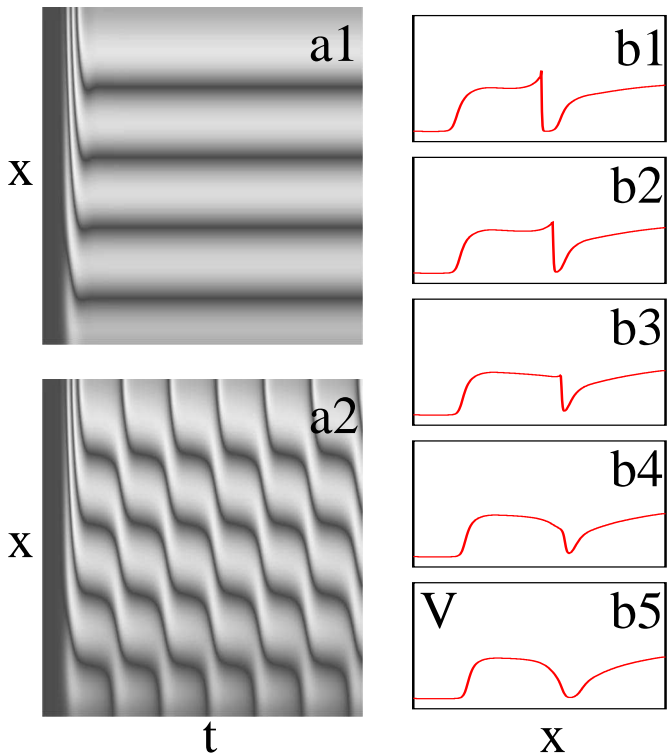}
\caption{\label{alternansfig}a1 and a2: 
Grey scale time-space plot of the alternans amplitude $|a(x,m)|$ with 
black corresponding to $a(x,m)=0$ and white corresponding to large
values of the amplitude.
The pacing period in both figures is 162 ms and the number of pacing grid
points is $n=6$ in a1 and $n=10$ in a2. In a1, one can see discordant
alternans where the nodes, regions of zero amplitude (and thus 
in black), are
not moving. In a2, the discordant alternans leads to nodes
that are traveling towards the pacing site. 
b1 to b5: successive profiles of V(x), taken at 20 ms intervals, during a
conduction block event. The waves are initiated on the left hand side
of the cable. In b1, one can see an action potential propagating from
the left towards the right with its typical sharp wavefront. The
smoother waveback of the wave due to the previous stimulus is on the
right hand side of the plot. The wavefront propagates faster than the
wave back (b1 to b3) and finally reaches a region that has not yet
recovered and can not be rapidly activated. Reaching this region (b4),
the wave comes to a stop and fails to propagate further (this is called
throughout this paper a conduction-block). In b5, one can see that the
sharp wavefront has disappeared. The cable is 10 cm long and V
ranges from -100 mV  to 100 mV.} 
\end{figure}

Let us first consider a one dimensional cable. For small enough 
pacing period,  
a period doubling instability to alternans takes place.
After a transient regime where the amplitude of alternans grows, a
steady state  is reached. This regime is characterized by the fact
that each stimulus will elicit an action potential all along the cable
and that the action potentials elicited by two successive stimulus at
a given point in space will have different durations (a short action
potential follows a long one  which follows a short one and so on)\cite{LewGue90}. 
This allows us to define, at a given
point in space $x$, the amplitude of alternans, $a(x,m)$,
for the  $m^{th}$ stimulus as  the  difference
between the APD due to this  stimulus and
the APD due to the
previous stimulus 
multiplied by $(-1)^m$ so that $a$ 
does not change sign every stimulus.
In the case of concordant alternans,  $a$ 
is roughly constant in space and
does not change sign. 
In our simulations, however, since the size of the simulated cable is
bigger than the characteristic wavelength of alternans
\cite{EchKar02}, we  observe discordant alternans. In this case
$a(x,m)$ is not constant in space and changes
sign when going along the cable.
For example, if a point in the cable with positive
$a$ exhibits a sequence of beats that is long
short long short... then a point of the cable with negative $a$
will exhibit a sequence of beats that is short long short long....
Consequently, between regions of   
opposite $a$, there is a point where $a=0$,
corresponding to an alternans node. 

In the case of discordant alternans, one can observe two distinct
regimes (see  \cite{EchKar02} for a full theoretical analysis) that
are characterized by the spatio-temporal evolution of $a$. The
first regime, called standing alternans, is characterized by stationary
alternans nodes. This is shown
shown in Fig. \ref{alternansfig} a1, where we show a gray-scale
space-time plot of $a$.
In contrast, the alternans nodes in the second regime, 
called traveling wave alternans, are non-stationary and are
traveling towards the pacing site.
This is illustrated in Fig \ref{alternansfig} a2 which 
shows that $a$ is changing slowly in time with a given period.
In addition, in this regime, $a=0$  at the pacing
site and $|a|$ increases when moving away from the pacing site.

A change in the dynamics of the alternans nodes was 
observed earlier in the theoretical study of Echebarria
and Karma \cite{EchKar02}. There, this change was accomplished
by  changing the properties of the tissue (i.e. by
changing the model).
Here, in contrast,
the only difference between the simulations shown in Fig
\ref{alternansfig} a1 and a2 is the size of the pacing region:
$n=6$ in a1 and $n=10$ in a2. 
This finding already shows that the
size of the pacing region may play a critical role in the alternans
induced arrhythmias. 
A rationale for
the effect of changing the size of the 
pacing region  will be given later in the manuscript (end of Sec. V).

\begin{figure}
\includegraphics[width=0.5\textwidth]{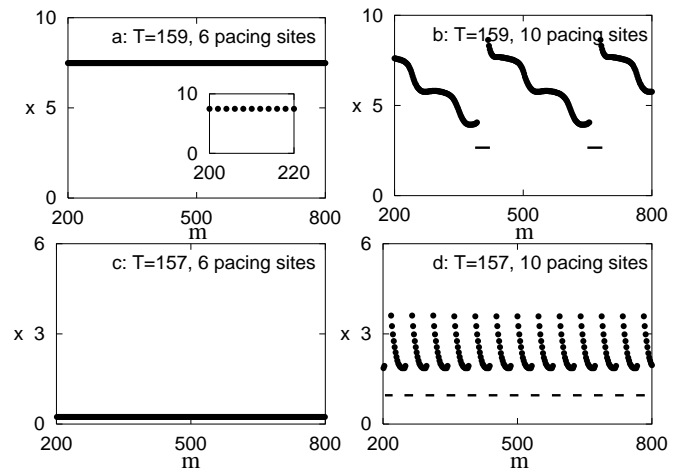}
\caption{Position of the conduction block as a function of the
stimulus number m for different pacing period $T$ (in ms) and pacing region
sizes $n$. In a, the inset shows that the
conduction block leads to a 2:1 rhythm. 
In both b and d, the conduction block travels towards the pacing site
during a number of stimuli, followed by a small number of stimuli
during which there is no conduction block. In b and d the thick lines correspond to the stimuli where no conduction block is observed.
One should note the similarity between the trajectories of 
conduction blocks for $T=159$ ms and $n=10$ (resp. $n=6$) 
and the trajectories of the alternans nodes for  $T=162$ ms  and $n=10$ 
(resp. $n=6$) shown in figure \ref{alternansfig} a1 and a2.
\label{figwaves}}
\end{figure}

Decreasing the stimulation period further leads to conduction blocks. 
In other words, some stimuli are not able to create an action potential 
that propagates all along the cable. 
Two different types of 
conduction blocks were observed: 
1) {\it conduction block at the pacing site}; 
the stimulus is not able to elicit an action potential at all 
and 2) 
{\it conduction block away from the pacing site}; 
the stimulus creates an action potential that begins to propagate 
along the cable and this action potential fails to propagate at a given point 
in space away from the pacing site. The latter type of 
conduction block is illustrated in Fig. \ref{alternansfig} b1-5. 
Hence, a conduction block can be characterized  by the point in space 
where it took place and also by the index of the stimulus 
that created the action potential which failed to propagate at this point. 

In our simulations, we found that the dynamics of 
the conduction blocks depends on  
both the size of the paced region and on the pacing period. 
Examples of the observed dynamics of the conduction blocks
are illustrated in Fig.  \ref{figwaves}, where
we plot the position of the conduction block as a function of the 
pacing cycle number m. 
In fig. \ref{figwaves}a and c, the block occurs at a fixed 
location (either away from the pacing site or at the pacing site) 
once every other stimulus (i.e. a 2:1 rhythm). 
Examples of more complex behavior are illustrated in Fig. 
\ref{figwaves}b and d
where the location of the conduction block 
is non-stationary and forms a periodic pattern.  
In these cases, a conduction block forms at a certain location
away from the pacing site.
In subsequent beats, the position of this conduction block moves 
closer to the pacing site 
(during this regime every other stimulus is blocked).
Eventually, the block disappears, after which no conduction 
blocks are observed during several pacing cycles 
(i.e. propagation with a discordant alternans pattern). 
Then, the conduction block reforms at precisely the same 
location as previously and the sequence restarts. 

This gives rise to a 
periodic pattern of conduction blocks with a certain ``amplitude",
defined here as the
distance between the site where the conduction block first forms and 
the site where the conduction blocks disappears. 
We found that this amplitude can take on values between 2 cm
(see Fig. \ref{figwaves}d) and 5 cm  (see Fig. \ref{figwaves}b) 
and that for a given pacing period, both the amplitude of the conduction 
block wave and its period were dependent on the number of pacing grid
points. 
We also found that the trajectory of the conduction blocks in Fig. 
\ref{figwaves}(b), which is close to the onset of conduction blocks,
is strikingly similar to the  dynamic pattern of the location of the
alternans nodes shown in Fig.  \ref{alternansfig} a2.
These alternans nodes also display periodic movement towards the pacing site
with slow movement followed by fast movement. 
This movement is presumably due 
to the use of zero flux boundary conditions which 
favors the existence of an extremum at the boundaries:
slow movement corresponds to an extremum of the alternans amplitude 
near the boundary  while fast movement corresponds to an alternans
node near the boundary. 
A phase diagram representing the nature of conduction blocks as a
function of the number of pacing grid points and of the period is shown in
Fig. \ref{phasediag}. It is also  worth mentioning that the use of slightly
irregular pacing times, modeled via the inclusion of a noise term with
a variance of 1 ms, did not change significantly the phase diagram. 

\section{Results in a sheet of cardiac tissue}
 
 From the above, it should now be clear how a slight inhomogeneity in 
the size of the pacing area can lead to reentry. 
Envision a 2D sheet of cardiac tissue 
(modeled by Eq.\ref{oned} using two space variables instead of one), 
paced from one side
by a domain containing two widths, $n$ for the upper part
and $n^*$ for the lower part, as shown in Fig. \ref{2d}. 
Neglecting spatial coupling  in the direction perpendicular
of the wave propagation, the lower and upper part of the tissue
will see waves originating from domains with different sizes. 
Reentry should be possible when 
the spatial locations of the conduction blocks in the two 
domains are significantly different.
Then, the electrical wave will be blocked in one
part of the tissue while propagating normally in the other part, 
leading to the excitation of the tissue behind the conduction 
block  and to reentry.

\begin{figure}
\begin{center}
\includegraphics[width=0.4\textwidth]{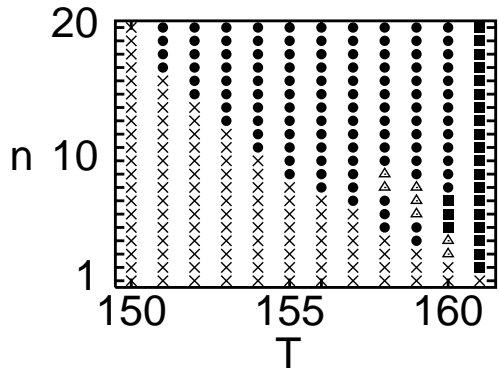}
\caption{Phase diagram showing the type of conduction block:
filled circles represent periodic conduction block waves, crosses
represent stationary conduction blocks at the pacing site, 
triangles
represent stationary conduction blocks away from the pacing 
site and filled boxes represent no conduction block (i.e. discordant
or concordant alternans.). T is expressed in ms.
\label{phasediag}} 
\end{center}
\end{figure}

\begin{figure}
\begin{center}
\includegraphics[width=0.4\textwidth]{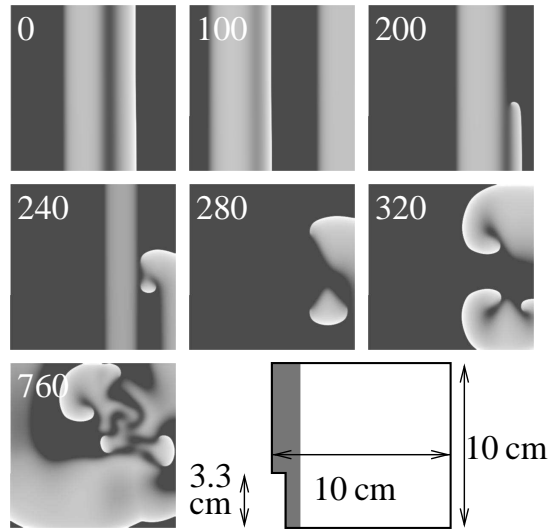}
\caption{ Series of gray-scale plots showing the 
the birth of a spiral wave. Time (shown in ms) is arbitrarily set 
to 0 at the start of the last stimulus. 
The gray-scale represents the membrane potential values with
white corresponding to maximum depolarization and black
corresponding to maximum repolarization.
At t=0 ms, one can see the wave elicited by the previous stimulus 
propagating from left to right. Because of the discordant alternans, 
the APD differs strongly across the sheet with 
the left most clear stripe corresponding to a region where 
the APD is long (it is not a propagating wave despite the cells in this region are depolarized\cite{Bik02}) and the central dark stripe corresponding 
to a region where the APD is short.  
At t=100 ms, one can see on the left hand side of the plot 
the wave elicited by the next stimulus while on the right hand side 
one can see the waveback of the wave seen in the t=0 ms frame.
At t=200 ms, a conduction block occurred in the top part of 
the sheet while in the bottom part the wave was able to continue
to propagate. 
This partial wave block results in a broken wave front which leads to
the birth of reentry as  
can be seen in the t=240 ms, 280 ms and 320 ms frames. 
As this wave front curls and attempts to reenter previously 
excited tissue with a long repolarization time (gray stripe in 
t=240 ms frame), it breaks up into two spiral tips (t=280 ms). 
Finally at t=760 ms, one can see further instabilities have led
to  a disordered activity similar to ventricular fibrillation. 
The geometry of the computational domain is also shown 
schematically where the gray region, exaggerated for clarity,
corresponds to the cells  that are paced.
The width of the 
bottom part  of this region is
$n^*=8$ and of the top part $n=9$ grid points. 
}
\label{2d}
\end{center}
\end{figure}

We can estimate the likelihood of reentry in 2D, paced by 
domains of size $n$ and $n^*$, 
based on our 1D results.
Reentry is likely when (i) 
a cable paced  with $n$
shows a conduction block while a cable paced with $n^*$ does not
exhibit a conduction block, or vice-versa,
or (ii) the difference in 
spatial location between the 
conduction blocks in cables paced with $n$ and 
$n^*$ becomes large. 
We found that for $n^*=n\pm 1$, 
one or both of these conditions were
met in most of the phase space for which conduction blocks occur
away from the pacing site (i.e. filled circles in 
Fig. \ref{phasediag}).
In most cases, 
the conditions were met within a few hundred stimuli while
occasionally  a larger number of stimuli was necessary. 

Of course, the preceding arguments neglect spatial coupling
perpendicular to the wave propagation direction. 
Nevertheless, we found that the phase diagram was a 
valuable predictor  for the occurrence of reentry in 
full scale 2D simulations.
A typical example of our 2D simulations,
performed using the same 
numerical scheme as in our one dimensional simulations,
is displayed in 
Fig. \ref{2d}, where we show a series of gray-scale plots of the 
membrane potential, with black corresponding to repolarized tissue 
and white corresponding to depolarized tissue.
In this example, the width of the two pacing domains 
differed by a single grid point: $n^*=8$ vs. $n=9$.
Both domains are paced with a constant period of 158 ms, 
leading to traveling conduction blocks 
with slightly different periodicity.
During the
roughly 100 first stimuli, the position of the conduction block 
was nearly identical in the upper and lower parts of the tissue.
During the subsequent stimulus, however, 
a wave block remains in the upper part of the tissue but {\it not} in
the lower part of the tissue, which now allows unblocked propagation
(see Fig. \ref{figwaves}b and d).
As time progresses, the propagating wave in the 
lower part  reenters the upper part behind the wave block, 
leading to a spiral wave. For clarity, we have stopped stimulating the tissue 
once the reentry appeared, although we 
have verified that 
continuous stimulation also resulted in long-lasting reentry.
In addition, we have checked that 
smaller domains were also able to produce reentry and that
using a line of 
stimulation with a thickness varying randomly can also lead to 
sustained reentry.

\section{Coupled map model}

The results from the full ionic model can be reproduced
by a coupled map model similar to the ones used in 
previous work \cite{Foxetala02, EchKar02}.
To describe the dynamics of alternans nodes and conduction blocks, we
consider the following two fields: $T_{up}(x,n)$, which is the time at
which an action potential is elicited in $x$ due to the $n^{th}$
stimulus and $T_{rep}(x,n)$, which is the time at which it ends. 
The equation for the repolarization time $T_{rep}$ is the same 
for both the cable and the pacing domain:
\begin{eqnarray}
T_{rep}(x,n)&=&T_{up}(x,n)+
APD(DI(x,n))+\nonumber\\
&+&\xi^2 \nabla^2 T_{rep}(x,n)+w\partial_x T_{rep}(x,n)\label{rep_eq}
\end{eqnarray}
The first two terms in this equation
express the fact that a single cell repolarizes
at time $T_{up}$ plus the duration of the action potential. 
The latter is taken from the restitution curve $APD(DI)$.
For the paced domain, this curve is calculated 
using a single cell while for the cable
it is calculated  1 cm away from the pacing domain, taken to be large.
In this region of the cable, we found that the 
restitution curve was minimally affected by the proximity of the 
pacing domain and that the 
dispersion of repolarization due to alternans was minimal. 
The two restitution curves are shown in Fig. \ref{coupledmap}d. 
The last two terms take into account spatial effects:
the diffusion
term indicates the perturbation due to inter-cellular coupling 
(we consider here that the repolarization wave is a phase wave and 
does not correspond to the propagation of a wave back)
and the last term  expresses the asymmetry introduced
by the pacing at one end of the cable \cite{EchKar02}.  
The length scales $\xi$ and
$w$  were chosen here to reproduce the phase-diagram 
of Fig. \ref{phasediag} in a semi-quantitative fashion. 

The equation describing $T_{up}$  in the pacing domain
is  $T_{up}(x,n)=T_n$, with $T_n$ being the pacing period,
while in the cable it is given by:
\begin{equation}
T_{up}(x,n)=T_n+\int_0^x dx' \frac{1}{c(DI(x',n))}\label{up_eq}
\end{equation}
where $c$ is the propagation speed of the wave front. 
This propagation speed
is a function of the $DI$, which itself is 
of course a function of $x$ and is the  difference between 
the arrival of the  $n^{th}$ action potential at $x$ and the 
repolarization following the $(n-1)^{th}$ stimulus:
$DI(x,n)=T_{up}(x,n)-T_{rep}(x,n-1)$.

The restitution curve also determined the occurrence of a conduction
block: 
we defined a conduction block to take place when the
restitution curve had no value for the diastolic interval computed
from Eq. \ref{up_eq}. 
In this case, we computed the repolarization
times between the origin and the position of the conduction block
using Eq. \ref{rep_eq} along with 
boundary condition $T_{rep}=T_{up}$ at the site of the conduction 
block. The repolarization time for the 
remainder of the cable (i.e. behind the 
conduction block) was set
to the repolarization time computed at the previous stimulus, except
in a small transition region,  in size equal to 
$\xi$, immediately behind 
the conduction block. There, a simple sigmoidal 
function was used that smoothly connected the 
repolarization times on both sides of this region. 
 
The reduced model is able to capture the essential features
of the full ionic model, including the striking dependence on the
size of the pacing region for both the alternans nodes
and the conduction blocks dynamics.  
The latter is shown in Fig. \ref{coupledmap}a-c 
where we plot the position of 
the conduction block as a function of pacing cycle for $w=0.025$ and 
$\xi^2=0.04$ which are comparable to the values presented in 
Table I of Ref. \cite{EchKar02} when considering a full ionic model.
All qualitative features of Fig. {\ref{figwaves}  can be 
readily recognized, including traveling conduction blocks with a
large amplitude (a)
\footnote{In this case, one should note that while in Fig \ref{figwaves} 
three plateaus of slowly moving conduction blocks are present, 
in Fig. \ref{coupledmap} a there are only two plateaus. },  
traveling conduction blocks with a
small amplitude (c), and stationary conduction block away from the 
pacing domain (b). 
Not shown here, but also found within the reduced
model, is 2:1 conduction block at the pacing site. 
For the values of $w$ and $\xi$ employed here, 
the boundaries between the different types of conduction block
in the phase diagram derived from the full model and the coupled-map 
model differed by at most 5 ms.  
\begin{figure}
\begin{center}
\includegraphics[width=0.4\textwidth]{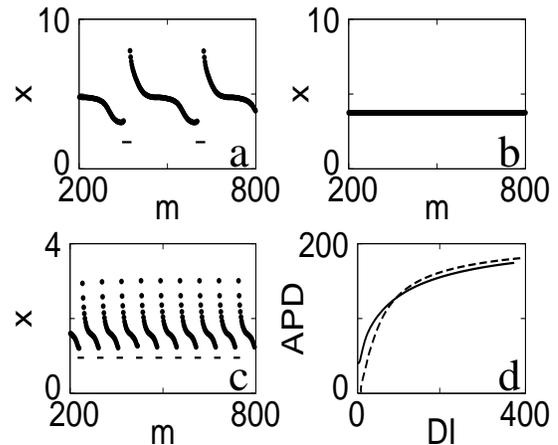}
\caption{
a-c: Position of the conduction block as a function of the
stimulus for different pacing period 
and pacing region sizes calculated using the coupled-map model (a:
T=161 ms, n=20, b: T=161 ms, n=10, c: T=156 ms, n=12). In a and c the thick lines correspond to the stimuli where no conduction block is observed.
d: The restitution curve used in the model
for the single element (solid line) and 
for the cable (dashed line). Note that the slope of the 
single cell curve is smaller than the slope of the whole cable curve.
} 
\label{coupledmap}
\end{center}
\end{figure}

Both the full ionic model and coupled map model 
show, for certain pacing periods, a transition  from stationary
conduction blocks to traveling conduction blocks as the 
size of the pacing domain is increased. 
Interestingly, a similar transition also occurs in the dynamics
of alternans nodes when one increases this size.
To understand the effect of changing the size of 
the pacing region on the nature of alternans
we consider the equation for the amplitude $a$ of the 
alternans derived in \cite{EchKar02}:
\begin{equation}
 \partial_t a=\sigma a -\int_0^x dx'\frac{a}{\Lambda}-
\tilde{w}\partial_x a+\tilde{\xi}^2\partial_{xx} a-ga^3,\label{ampek}
 \end{equation}
Here, $\sigma a$ represents a linear growth term, 
$ga^3$ is a nonlinear restabilizing term and 
$\Lambda$, $\tilde{w}$ and $\tilde{\xi}$ are length scales which  
are \textit{a priori} different from the ones $w$ and $\xi$  
used in Eqs. \ref{rep_eq}, \ref{up_eq}.
The transition between traveling nodes in our simulations (large
pacing domains) and standing nodes (smaller pacing domains) 
can be understood when 
realizing the size of the pacing domain dictates the 
boundary condition at $x=0$ of Eq. \ref{ampek}. 
Pacing a cable using a large piece of tissue results in a pacing 
domain that essentially behaves as a single isolated cell.
If the restitution curve of a single cell (solid line in  Fig.
\ref{coupledmap}b) has a slope less than one,
this domain will not display
alternans and the amplitude equation needs to be solved with 
boundary condition $a=0$.
For a smaller pacing domain, however, spatial coupling 
becomes important and the relevant restitution curve
becomes steeper (dashed line in Fig. \ref{coupledmap}b), 
which can lead to alternans. 
Thus, for smaller domains
the relevant boundary condition for the amplitude equation
is $\partial_x a=0$.  
A standing wave solution to the amplitude equation can be  
described by $a=a_0 cos(2\pi \lambda x + \phi)$,  where 
$1/\lambda$ is the separation between the nodes.
This solution (with $a_0 \ne 0$) can only satisfy 
the $\partial_x a=0$ boundary condition and {\it not} the $a=0$
boundary condition. 
This is easy to see when one considers the standing wave solution 
in Eq.  \ref{ampek} 
at $x=0$: the only non zero term at the r.h.s. is $\partial_x a$
which requires $a_0 2\pi/\lambda =0$ and finally $a_0=0$.
Hence, a standing wave solution is only possible 
for small pacing sizes and  
increasing the size of the pacing domain suppresses 
the standing wave solution. 
It is likely that this qualitative change in the nature 
of allowed solutions underlies the transition between 
stationary to traveling blocks observed in the ionic and 
coupled map model.

\section{Discussion}

To conclude we have found that the spatio-temporal structure  of
alternans and the dynamics of conduction blocks is
strongly influenced by both the size of the paced domain and the 
pacing period. This dependence provides a novel arrhythmogenic mechanism
which we illustrate in a  homogeneous two dimensional sheet of tissue, 
paced by two domains that vary slightly in size. 
The reentry is initiated through pacing with a constant period and 
hence does not require an abruptly changing pacing frequency 
nor a symmetry-breaking change of
location as in previous studies \cite{Quetal00a}.
We have found, 
using different ionic models, including a detailed canine model
\cite{Foxetala02} and the simplified three variable Fenton-Karma model 
\cite{Fenetal02,FenKar98}, that the results we present here are 
partially model-dependent. 
This is perhaps not surprising, as other arrhythmogenic mechanisms have
been show to depend on the details of the electrophysiological 
model \cite{Rap01}.
Nevertheless, a reduced coupled map model
was able to reproduce, both  qualitatively and, to some degree, 
quantitatively, the results presented here. Consequently, we expect that the 
dynamics of conduction blocks will 
depend critically on the pacing protocol in 
a wide range of detailed models.

Of course, the ultimate test should come in the form of 
experimental studies similar to the one conducted by Fox \textit{et al} \cite{Foxetala02}   but with a varying in size stimulation region.
It should be possible to conduct quantitative
studies of Purkinje fibers, which 
conduct the electrical stimulus in an
actual heart to the ventricles and which
penetrate the heart wall to varying depths. 
These fibers can be isolated, resulting in linear strands of cardiac
tissue \cite{Foxetala02}.
On the theoretical side, a further extension of this work will be
the formulation of
a coupled map model which takes into account the effects of
coupling transversely to the propagation direction of the wave.

\acknowledgements

The authors would like to thank
Flavio Fenton and Jeffrey Fox for useful discussions.  
They also would like to thank 
Alain Karma for  a critical reading of this manuscript.
Computations were performed in part on the National Science
Foundation Terascale Computing System at the Pittsburgh Supercomputing
Center. We acknowledge the National Biomedical Computation Resource
(NIH P41 RR-08605).
This research was supported in part by 
the NSF-sponsored Center for
Theoretical Biological Physics  (grant numbers PHY-0216576 and
0225630) (HH and WJR), and by NIH grant HL075515-01 (WJR).
\appendix

\section{Description of the model \label{LR1Dyn}}

The Luo-Rudy model \cite{LuoRud91}, and its subsequent refinements
(see e.g.  \cite{LuoRud94a}), have been
widely used either in their original form or in modified forms in
numerical studies of wave propagation in cardiac
tissue \cite{Quetal00b}. 
The model describes the voltage and time dependence of the ionic currents 
$\Sigma_{ion} I_{ion}$
used in equation for the transmembrane
potential $V$ of a single cell: 
\begin{eqnarray}
C \frac{d V}{dt}=- \Sigma_{ion} I_{ion}
\end{eqnarray}
where C is the membrane capacitance. 
In the original Luo-Rudy model, the total ionic current is given as
\begin{eqnarray}
\Sigma_{ion} I_{ion}
 =-(I_{Na}+I_{si}+I_K+I_{K1}+I_{Kp}+I_{b})
\end{eqnarray}
where $I_{Na}=G_{Na}m^3hj(V-E_{Na})$ is the fast sodium current, 
$I_{si}=G_{si}df(V-E_{si})$ is the slow inward current representing
the L-type calcium current,
$I_{K}=G_{K}xx_1(V-E_{K})$ is the time dependent potassium current, 
$I_{K1}=G_{K1}K1_{\infty}(V-E_{K1})$ is the time independent potassium current, 
$I_{Kp}=G_{Kp}K_p(V-E_{Kp})$ is the plateau  potassium current and
$I_{b}=G_b(V-E_b)$ is the background current.
In these expression, $m,h,j,d,f$ and $x$ are gating variables 
describing the opening and closing of ionic channels  
and the dynamics of these variables is described by 
nonlinear ordinary differential
equations of the form:
\begin{eqnarray}
\frac{dy}{dt}&=& \frac{y_{\infty}(V)-y}{\tau_y(V)}
\end{eqnarray}
where $y$ represents one of the gating variable.
Finally, the equations for the currents are supplemented by
an expression for the calcium concentration:
\begin{eqnarray}
\frac{[Ca]_i}{dt}&=& -10^4 I_{si}+0.07(10^{-4} [Ca]_i)
\end{eqnarray}

Explicit expressions for the constants can be found in the original
work of Luo and Rudy (Ref. \cite{LuoRud91}).
Here, we have made two   modifications
to allow propagation of spiral waves in a small system.
First, we have reduced $G_{si}$ from 0.09 to 0.055.
Second, we have sped up the calcium dynamics by
altering the time scales of the $d$ and $f$ gates. Specifically, 
both $\tau_d$ and $\tau_f$ were multiplied by 0.8. 
 
 Initial conditions that were used consisted of a cable initially polarized that was then paced with decreasing intervals between stimuli until the desired period was reached. 

%\bibliography{heart}

\end{document}